\documentclass{PoS}
\usepackage{psfrag,epsfig,graphicx,graphics,xcolor}
\usepackage{wrapfig}
\usepackage{amsmath}

\newcommand{\be}{\begin{equation}}
\newcommand{\beq}{\begin{equation}}
\newcommand{\ee}{\end{equation}}
\newcommand{\eq}{\end{equation}}

\newcommand{\bea}{\begin{eqnarray}}
\newcommand{\eea}{\end{eqnarray}}

\newcommand{\veck}{{\bf k}}
\newcommand{\vecki}{{\bf k}_i}
\newcommand{\veckone}{{\bf k}_1}
\newcommand{\vecktwo}{{\bf k}_2}
\newcommand{\veckj}{{\bf k}_{J}}
\newcommand{\veckji}{{\bf k}_{J,i}}
\newcommand{\veckjone}{{\bf k}_{J,1}}
\newcommand{\veckjtwo}{{\bf k}_{J,2}}

\newcommand{\deins}[1]{{\rm d}#1\,}
\newcommand{\dzwei}[1]{{\rm d}^2#1\,}
\newcommand{\dk}{\dzwei{\veck}}

\newcommand{\dkone}{\dzwei{\veckone}}
\newcommand{\dktwo}{\dzwei{\vecktwo}}

\newcommand{\dsigma}{\deins{\sigma}}
\newcommand{\dsigmahat}{\deins{{\hat\sigma}_{\rm{ab}}}}
\newcommand{\dnu}{\deins{\nu}}

\newcommand{\dx}{\deins{x}}
\newcommand{\dxone}{\deins{x_1}}
\newcommand{\dxtwo}{\deins{x_2}}
\newcommand{\dyjetone}{\deins{y_{J,1}}}
\newcommand{\dyjettwo}{\deins{y_{J,2}}}
\newcommand{\dphij}{\deins{\phi_{J}}}
\newcommand{\dphijone}{\deins{\phi_{J,1}}}
\newcommand{\dphijtwo}{\deins{\phi_{J,2}}}
\newcommand{\dtwojets}{{\rm d}|\veckjone|\,{\rm d}|\veckjtwo|\,\dyjetone \dyjettwo}
\newcommand{\shat}{{\hat s}}

\newcommand{\chihat}{{\omega}}

\title{Mueller-Navelet jets at LHC: the first complete NLL BFKL study\thanks{ 
We thanks D.~Colferai and F.~Schwennsen for many discussions at the initial stage of this work, and M.~Fontannaz for fruitful exchanges. This work is partly supported by the Polish Grant NCN No
DEC-2011/01/D/ST2/02069, the French-Polish collaboration agreement 
Polonium, the P2IO consortium and the Joint
Research Activity "Study of Strongly Interacting Matter" (acronym 
HadronPhysics3, Grant Agreement n.283286) under
the Seventh Framework Programme of the European Community.}}

\ShortTitle{Mueller Navelet jets at LHC: the first complete NLL BFKL study}

\author{B.~Duclou\'e\\
        LPT, Universit\'e Paris-Sud, CNRS, 91405, Orsay, France\\
   E-mail: \email{Bertrand.Ducloue@th.u-psud.fr}}

\author{L. Szymanowski\\
       National Center for Nuclear Research (NCBJ), Warsaw, Poland\\
       E-mail: \email{Lech.Szymanowski@fuw.edu.pl}}

\author{\speaker{S. Wallon}
\\
       LPT, Universit\'e Paris-Sud, CNRS, 91405, Orsay, France\\
       UPMC Univ. Paris 06, facult\'e de physique, 4 place Jussieu, 75252 Paris Cedex 05,
France\\
       E-mail: \email{Samuel.Wallon@th.u-psud.fr}}

\abstract{Mueller Navelet jets were proposed 25 years ago as a decisive test of BFKL dynamics at hadron colliders. We here present the first next-to-leading BFKL study of the cross section and azimuthal decorrelation of these jets. This includes both next-to-leading corrections to the Green's function and next-to-leading corrections to the Mueller Navelet vertices. The obtained results for standard observables proposed for studies of Mueller Navelet jets show that both sources of corrections are of equal and big importance for final magnitude and behavior of observables, in particular for the LHC kinematics investigated here in detail. Our analysis reveals that the observables obtained within the complete next-to-leading order BFKL framework of the present work are quite similar to the same observables obtained
 within next-to-leading logarithm DGLAP type treatment. There is still a noticeable difference in both treatments for the ratio of the azimuthal angular moments $\langle \cos 2 \phi \rangle / \langle \cos \phi \rangle.
$}

\FullConference{Sixth International Conference on Quarks and Nuclear Physics,\\
		April 16-20, 2012\\
		Ecole Polytechnique, Palaiseau, France}

\begin{document}
\section{Introduction}
\label{Sec_Int}

The understanding of the high energy regime of QCD is one of the key questions of particle physics.
In the semi-hard regime of a scattering process in which $s \gg -t$,  logarithms of the type $[\alpha_s\ln(s/|t|)]^n$ have to be resummed, giving the 
leading logarithmic (LL) Balitsky-Fadin-Kuraev-Lipatov (BFKL)  Pomeron \cite{BFKL_LL} contribution to the gluon Green's function.
The question of testing such effects experimentally then appeared, and
 various  tests have been proposed  in inclusive \cite{test_inclusive}, semi-inclusive  \cite{test_semi_inclusive} and exclusive processes \cite{test_exclusive}. The basic idea is to select specific observables which reduce the importance of usual collinear logarithmic effects \`a la DGLAP \cite{DGLAP}  with respect to the BFKL one:
 the involved transverse scales should thus be of similar order of magnitude.
We here consider the Mueller Navelet (MN) jets \cite{Mueller:1986ey} in hadron-hadron colliders, defined as being separated by a large relative rapidity, while having two similar transverse energies.  
In a DGLAP scenario, an almost back-to-back emission is expected, while the allowed
 BFKL emission of partons between these two jets  leads in principle to a larger cross-section, with a reduced azimuthal correlation between them.
We report on recent results  where both the  NLL Green's function \cite{BFKL-NLL} and the NLL result for the jet vertices \cite{Bartels:vertex} are taken into account\footnote{These vertices have been recently recomputed in Ref.~\cite{Caporale:2011cc}, in a full BFKL approach.}. These new results have been obtained based on a fast Fortran code, which allowed us to go beyond the studies of Ref.~\cite{us_MN} where we developped an exploratory Mathematica code. Detailed results will be presented elsewhere~\cite{DSW}.

\section{NLL calculation}
\label{sec:NLLcalculation}


The two hadrons collide at a center of mass energy $\sqrt{s}$ producing two very forward jets, whose transverse momenta  are labeled by Euclidean two dimensional vectors $\veckjone$ and $\veckjtwo$, and by their azimuthal angles  $\phi_{J,1}$ and $\phi_{J,2}$. The jet rapidities  $y_{J,1}$ and $y_{J,2}$  are related to the longitudinal momentum fractions of the jets via $x_J = \frac{|\veckj|}{\sqrt{s}}e^{y_J}$. 
For large $x_{J,1}$ and $x_{J,2}$, collinear factorization leads~to
\begin{equation}
  \frac{\dsigma}{\dtwojets} = \sum_{{\rm a},{\rm b}} \int_0^1 \dxone \int_0^1 \dxtwo f_{\rm a}(x_1) f_{\rm b}(x_2) \frac{\dsigmahat}{\dtwojets},
\end{equation}
where $f_{\rm a,b}$ are the parton distribution functions~(PDFs) of a parton a (b) in the according proton.
The  resummation of logarithmically enhanced contributions 
are included 
through $k_T$-factorization:
\begin{equation}
  \frac{\dsigmahat}{\dtwojets} = \int \dphijone\dphijtwo\int\dkone\dktwo V_{\rm a}(-\veckone,x_1)\,G(\veckone,\vecktwo,\shat)\,V_{\rm b}(\vecktwo,x_2),\label{eq:bfklpartonic}
\end{equation}
where the BFKL Green's function $G$ depends on $\shat=x_1 x_2 s$. The jet vertices
 $V_{a,b}$ were  calculated at
 NLL order in Ref.~\cite{Bartels:vertex}.
Combining the PDFs with the jet vertices one writes
\begin{eqnarray}
  \frac{\dsigma}{\dtwojets} 
\!&=& \!\!\int \dphijone\dphijtwo \! \!\int \! \dkone\dktwo \Phi(\veckjone,x_{J,1},-\veckone)\,G(\veckone,\vecktwo,\shat)\,\Phi(\veckjtwo,x_{J,2},\vecktwo) \,,\nonumber
\\
\mbox{where }&& \quad \Phi(\veckji,x_{J,i},\vecki) = \int dx_i\, f(x_i)\, V(\vecki,x_i).
\end{eqnarray}
In view of the azimuthal decorrelation we want to investigate, we define the  coefficients
\beq
  \mathcal{C}_m \equiv \!\!\!\int \!\! \dphijone\dphijtwo\cos\big(m(\phi_{J,1}-\phi_{J,2}-\pi)\big)\!\!\!\int\!\!\dkone\dktwo \Phi(\veckjone,x_{J,1},-\veckone)G(\veckone,\vecktwo,\shat)\Phi(\veckjtwo,x_{J,2},\vecktwo) , \nonumber
\eq
from which one can easily obtain the differential cross section and the azimuthal decorrelation as
\begin{equation}
  \frac{\dsigma}{\dtwojets} = \mathcal{C}_0 \quad {\rm and} \quad 
  \langle\cos(m\varphi)\rangle \equiv \langle\cos\big(m(\phi_{J,1}-\phi_{J,2}-\pi)\big)\rangle = \frac{\mathcal{C}_m}{\mathcal{C}_0} .
\end{equation}
The important step is then to use
 the LL-BFKL eigenfunctions
\begin{equation}
  E_{n,\nu}(\veckone) = \frac{1}{\pi\sqrt{2}}\left(\veckone^2\right)^{i\nu-\frac{1}{2}}e^{in\phi_1}\,,
\label{def:eigenfunction}
\end{equation}
although they
 strictly speaking do not diagonalize the NLL BFKL kernel.
In the LL approximation, 
\begin{equation}
  \mathcal{C}_m = (4-3\delta_{m,0})\int \dnu C_{m,\nu}(|\veckjone|,x_{J,1})C^*_{m,\nu}(|\veckjtwo|,x_{J,2})\left(\frac{\shat}{s_0}\right)^{\chihat(m,\nu)}\,,
\label{eq:cm2}
\end{equation}
\begin{align}
 {\rm where} \quad C_{m,\nu}(|\veckj|,x_{J})
=& \int\dphij\dk \dx f(x) V(\veck,x)E_{m,\nu}(\veck)\cos(m\phi_J) \,, \label{eq:mastercnnu}
\end{align}
and
$
  \chihat(n,\nu) = N_c\alpha_s/\pi
\chi_0\left(|n|,\frac{1}{2}+i\nu\right) ,$ with
$\chi_0(n,\gamma) = 2\Psi(1)-\Psi\left(\gamma+\frac{n}{2}\right)-\Psi\left(1-\gamma+\frac{n}{2}\right)\,.
$
The master formulae of the LL calculation~(\ref{eq:cm2}, \ref{eq:mastercnnu}) will also be used for the NLL calculation, the eigenvalue now turning to an operator containing a $\nu$ derivative \cite{Vera:2006un, Vera:2007kn}, 
which acts on the impact 
factors and effectively leads to a contribution to the eigenvalue
 which depends on the impact factors.

At NLL, the jet vertices are intimately dependent on the jet algorithm~\cite{Bartels:vertex}. We here use the cone algorithm, with the cone parameter $R=0.5$\footnote{A detailed study \cite{DSW}, based on the work of Ref.~\cite{Ivanov:2012ms} where the jet vertices were computed in an approximated small $R$ treatment, shows that the difference between an exact treatment and this approximation is small.}.
At NLL,
one should also pay attention to the choice of scale $s_0$. We find the choice of
scale  $s_0 =\sqrt{s_{0,1} \, s_{0,2}}$  with 
$s_{0,i}= \frac{x_{i}^2}{x_{J,i}^2}\veckji^2$ rather natural, since it does not depend on the momenta $\veck_{1,2}$ to be integrated out. Besides, the dependence with respect to $s_0$ of the whole amplitude can be studied,
when taking into account the fact that both the NLL BFKL Green's function and the vertex functions are $s_0$ dependent.
In order to study the effect of possible collinear improvement \cite{resummed},
we have, additionally, implemented for $n=0$ the scheme 3 of the first paper of Ref.~\cite{resummed}. This is only required by the Green's function since we could show by a numerical study that the 
jet vertices are free of $\gamma$ poles and thus do not call for any collinear improvement.
In practice, the use of Eqs.~(\ref{eq:cm2}, \ref{eq:mastercnnu}) leads 
to the possibility to calculate for a limited number of $m$ the coefficients $C_{m,\nu}$ as universal grids in $\nu$, instead of using a two-dimensional grid in $\veck$ space.
We use MSTW 2008 PDFs \cite{Martin:2009iq} and a two-loop strong coupling with a scale $\mu_R= \sqrt{|\veckjone|\cdot |\veckjtwo|}\,.$

\section{Results}
\label{sec:Results}

\begin{figure}
  \psfrag{sigma}{\scalebox{0.9}{$\sigma \left[ \frac{\rm nb}{{\rm GeV}^2} \right] $}}
  \psfrag{Y}{\scalebox{0.9}{$Y$}}

\begin{tabular}{cc}
\hspace{0cm}\includegraphics[width=7cm]{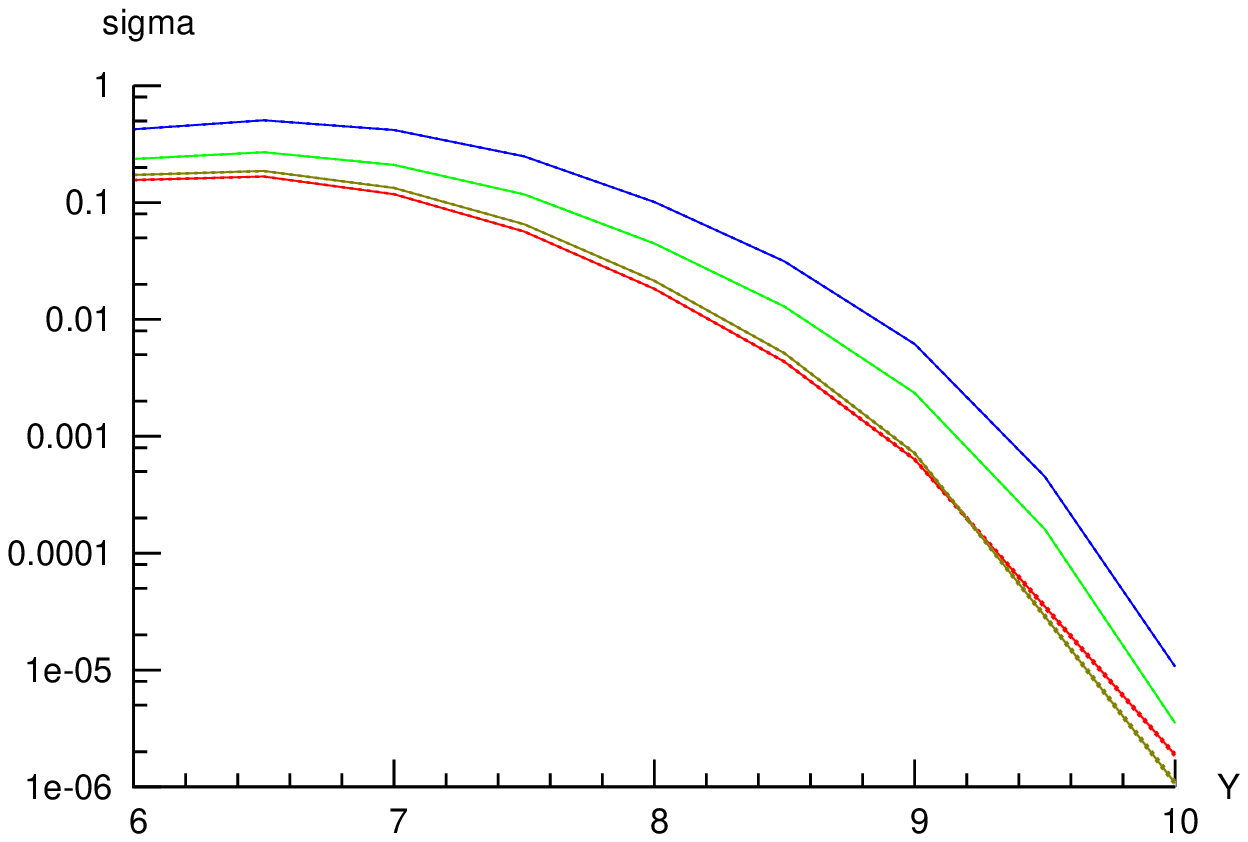} &  \psfrag{deltasigma}{$\frac{\Delta \sigma}{\sigma}$}
  \psfrag{s0}{}\psfrag{cubaerror}{}\psfrag{pdfset}{}\psfrag{mu}{}
  \psfrag{deltaC0}{$\delta\mathcal{C}_0 \left[\frac{\rm nb}{{\rm GeV}^2}\right] $}
  \psfrag{Y}{$Y$}
\includegraphics[width=7cm]{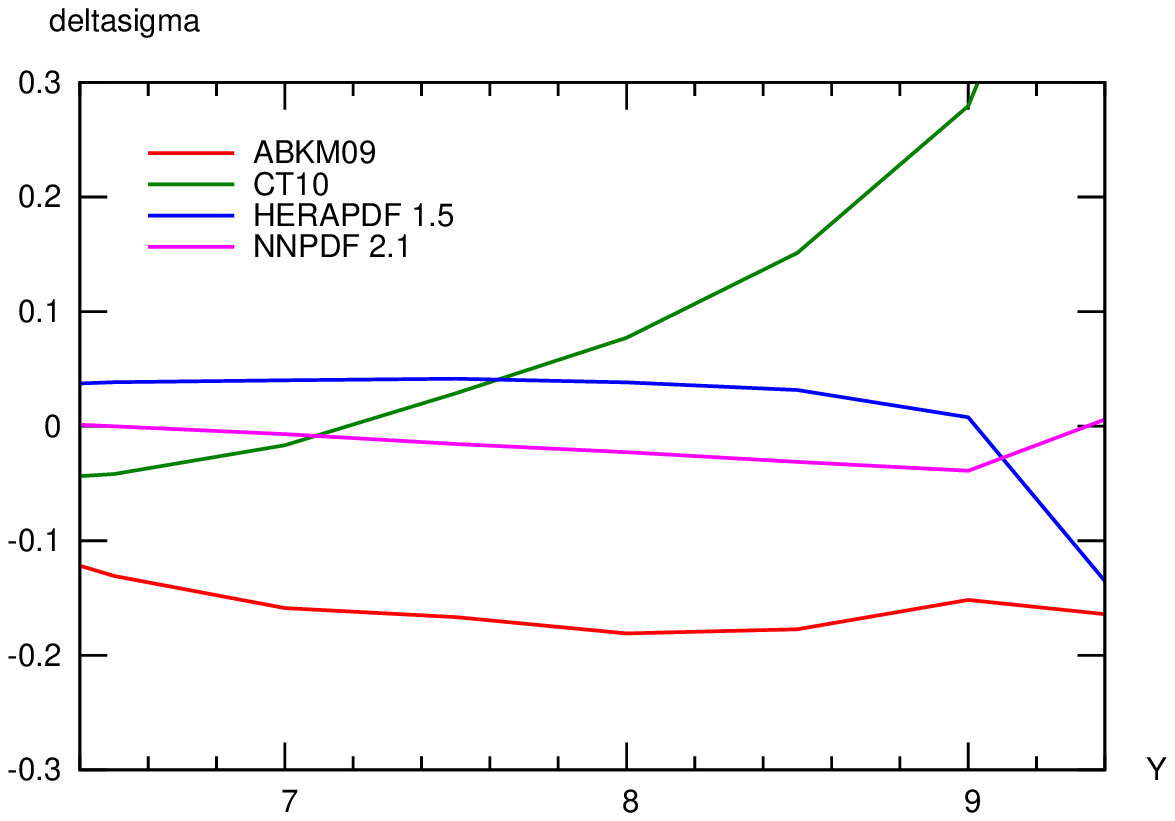}
\end{tabular}
\caption{Left: Differential cross section in dependence on $Y$ for $|\veckjone|=|\veckjtwo|=35\,{\rm GeV}\,,$ at $\sqrt{s}=7$~TeV.
Blue: pure LL result; Brown: pure NLL result; Green: combination of LL vertices with the collinear improved NLL Green's function; Red: full NLL vertices with the collinear improved NLL Green's function. Right: Relative variation of the cross-section in the full NLL approach, for various choices of PDFs with respect to MSTW 2008 ones.
}
\label{Fig:sigma}
\end{figure}

Fig.~\ref{Fig:sigma} (left)
displays the cross-section as a function of the relative jet rapidity $Y$, for the LHC  center of mass energy $\sqrt{s}=7\,{\rm TeV}$, for which most of LHC data are taken at the moment, while Fig.~\ref{Fig:sigma} (right) shows the relative variation of the cross-section with respect to MSTW 2008 when changing the PDFs according 
to Ref.~\cite{LHAPDF}. 
This explicitely shows 
the dramatic effect of the NLL vertex corrections, of the same order as the one for the Green's function \cite{Vera:2007kn,Marquet:2007xx}.

\begin{figure}
\begin{tabular}{ll}
 \psfrag{cos}{\raisebox{.1cm}{\scalebox{0.9}{$\langle \cos \phi\rangle$}}}
 \psfrag{Y}{\scalebox{0.9}{$Y$}}
\includegraphics[width=6.5cm]{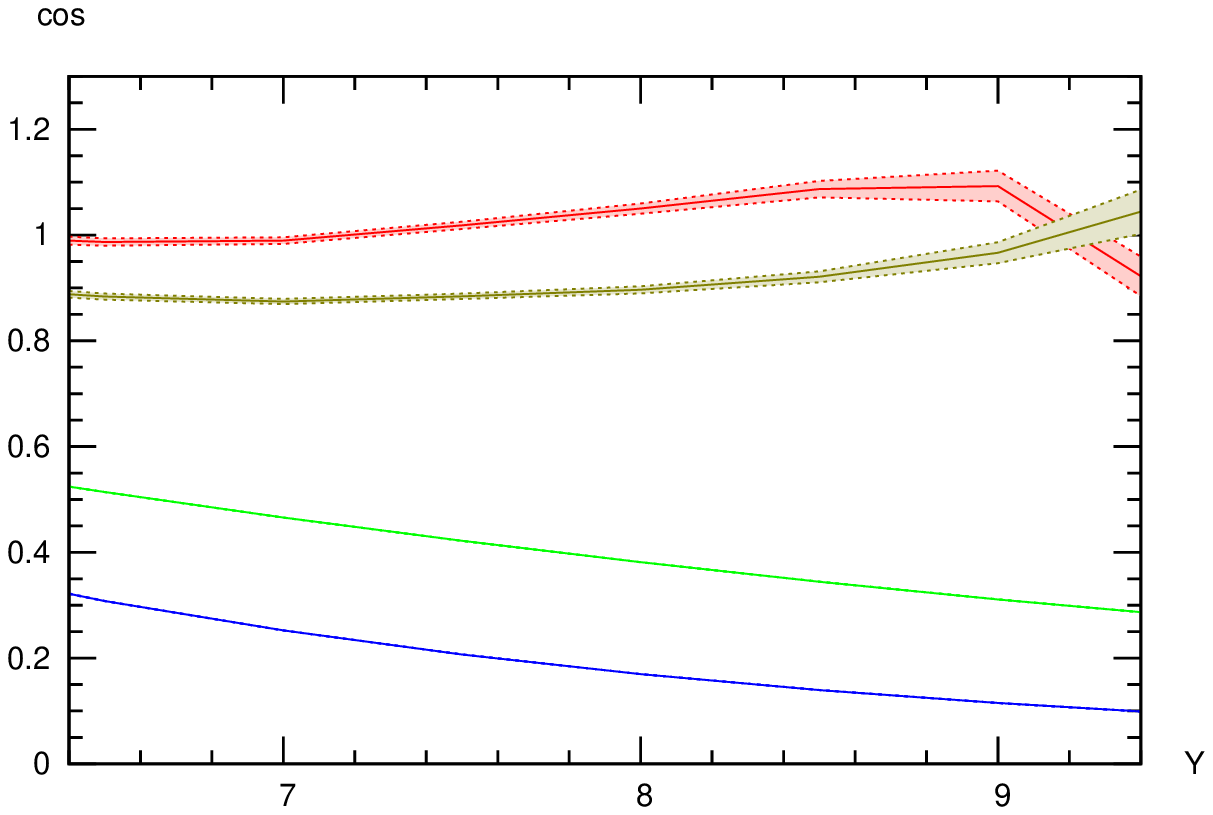} 
&
 \psfrag{deltacos}{$\frac{\Delta \langle \cos{\phi} \rangle}{\langle \cos{\phi} \rangle}$}
  \psfrag{s0}{}\psfrag{cubaerror}{}\psfrag{pdfset}{}\psfrag{mu}{}
  \psfrag{deltaC0}{$\delta\mathcal{C}_0 \left[\frac{\rm nb}{{\rm GeV}^2}\right] $}
  \psfrag{Y}{$Y$}
\hspace{1cm}\includegraphics[width=6.5cm]{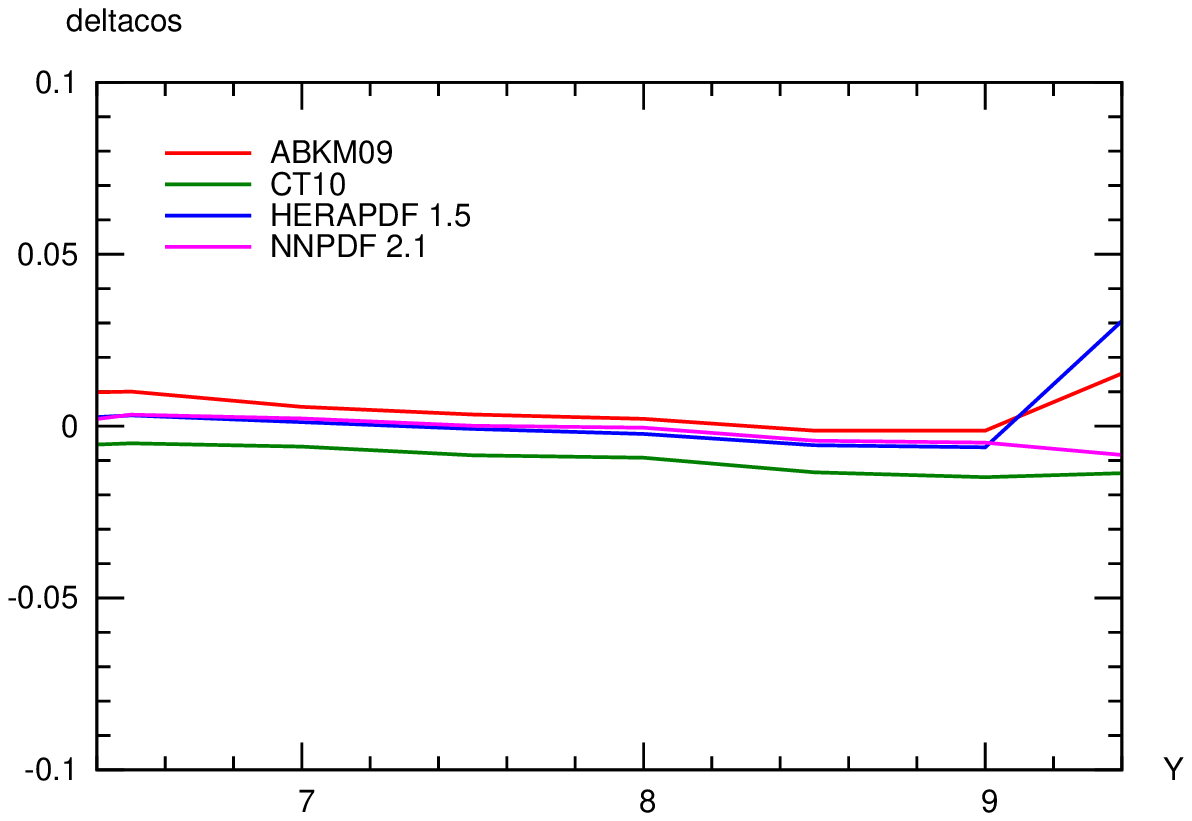}
\end{tabular}
\caption{Left: Azimuthal correlation in dependence on $Y$ for $|\veckjone|=|\veckjtwo|=35\,{\rm GeV}\,,$ at $\sqrt{s}=7$ TeV.
Blue: pure LL result; Brown: pure NLL result; Green: combination of LL vertices with the collinear improved NLL Green's function; Red: full NLL vertices with the collinear improved NLL Green's function. Right: Relative variation of the azimuthal correlation in the full NLL approach, for various choices of PDFs with respect to MSTW 2008 ones.
}
\label{Fig:correlation}
\end{figure}

 Fig.~\ref{Fig:correlation} (left)
displays the azimuthal correlation as a function of the relative jet rapidity $Y$, for the LHC  center of mass energy $\sqrt{s}=7\,{\rm TeV}$, while Fig.~\ref{Fig:correlation} (right) shows the relative variation of the cross-section with respect to MSTW 2008 when changing the PDFs using the sets of Ref.~\cite{LHAPDF}. The decorrelation based on our full NLL analysis is very small, similar to the one based on NLO DGLAP. Fig.~\ref{Fig:correlation} (right) shows that $\langle \cos \phi\rangle$ is much less sensitive to the PDFs than the cross section.

\begin{figure}
\begin{tabular}{cc}
 \psfrag{varied}{}
 \psfrag{cubaerror}{}
 \psfrag{cos}{\raisebox{.1cm}{\scalebox{0.9}{$\langle \cos 2\phi\rangle / \langle \cos \phi\rangle$}}}
 \psfrag{Y}{\scalebox{0.9}{$Y$}}
 \includegraphics[width=7cm]{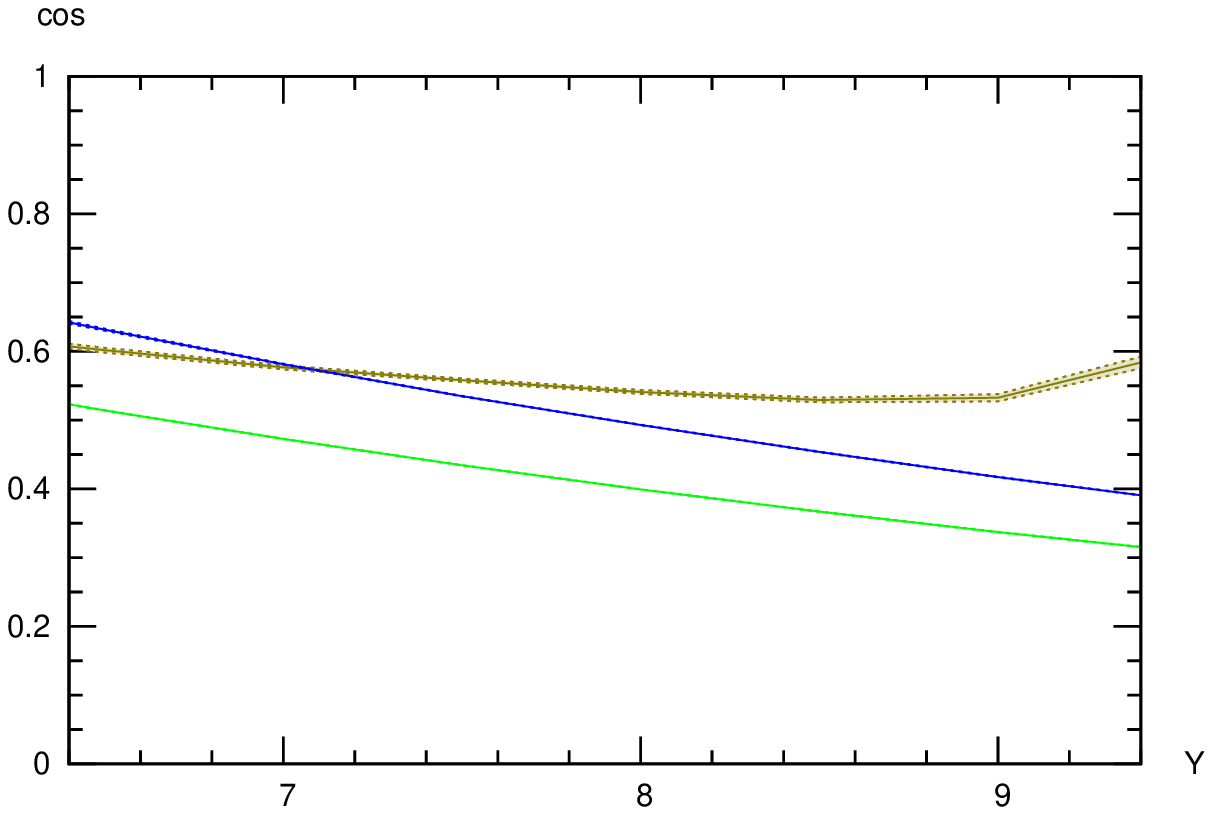} &
\psfrag{deltacos}{$\frac{\Delta ( \langle \cos{2 \phi} \rangle / \langle \cos{\phi} \rangle)}{\langle \cos{2 \phi} \rangle / \langle \cos{\phi} \rangle}$}
  \psfrag{varied}{}
  \psfrag{s0}{}
\psfrag{cubaerror}{}
\psfrag{pdfset}{}
\psfrag{mu}{}
  \psfrag{deltaC0}{$\delta\mathcal{C}_0 \left[\frac{\rm nb}{{\rm GeV}^2}\right] $}
  \psfrag{Y}{$Y$}
 \includegraphics[width=7cm]{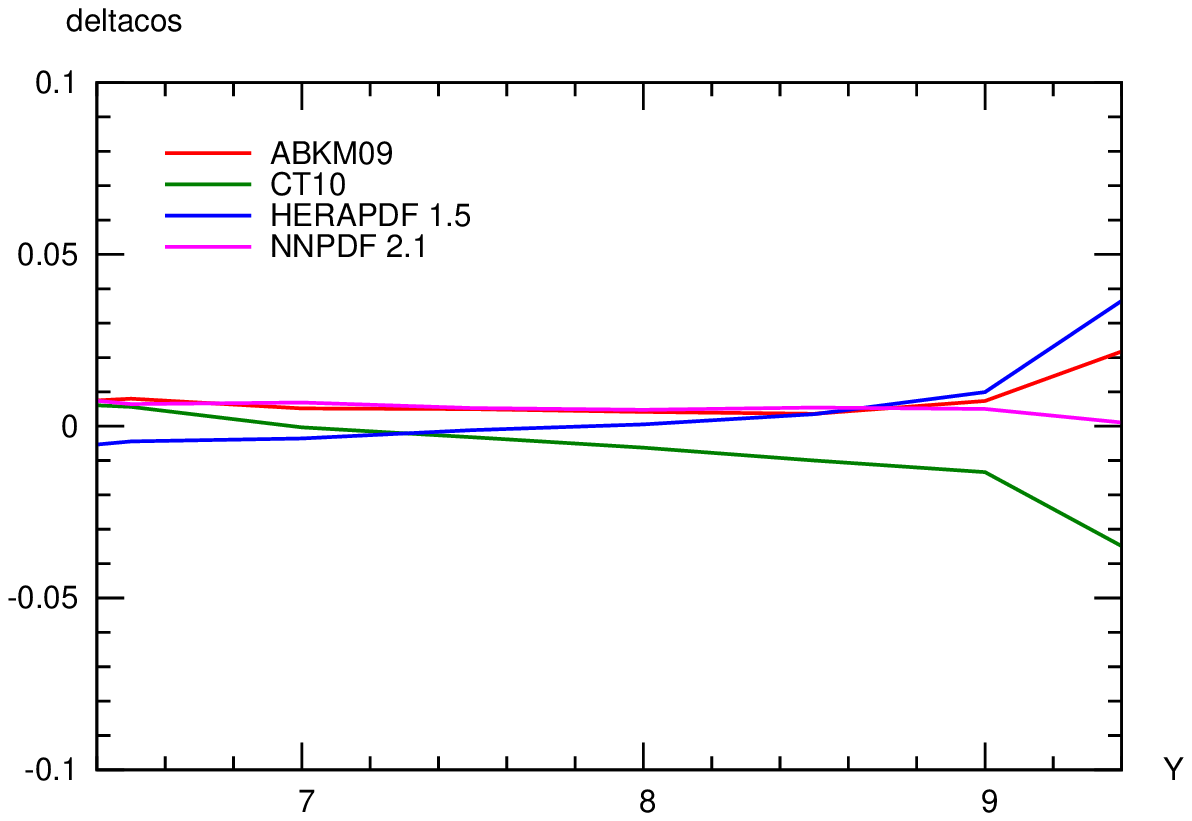}
\end{tabular}
\caption{Left: Ratio of azimuthal correlations $\langle \cos 2\phi\rangle / \langle \cos \phi\rangle$ in dependence on $Y$ for $|\veckjone|=|\veckjtwo|=35\,{\rm GeV}\,,$ at $\sqrt{s}=7$ TeV.
Blue: pure LL result; Brown: pure NLL result; Green: combination of LL vertices with the collinear improved NLL Green's function.
Right: Relative variation of $\frac{\langle \cos{2 \phi} \rangle}{\langle \cos{\phi} \rangle}$ in the full NLL approach when using other PDF sets than MSTW~2008.
}
\label{Fig:ratio_correlation}
\end{figure}

Detailed studies~\cite{us_MN} have shown that the main source of uncertainties is due to the renormalization scale $\mu_R$ and to the energy scale $\sqrt{s_0}$. \ This is particularly important for the azimuthal correlation, which, 
 when including a collinear improved Green's function, may exceed 1 for small $\mu_R=\mu_F$.
A noticeable difference can be expected between BFKL and DGLAP type of treatment for the ratio 
$\langle \cos 2\varphi\rangle / \langle \cos \varphi\rangle\,.$ The BFKL results are shown in Fig.~\ref{Fig:ratio_correlation}. We refer to Ref.~\cite{us_MN}
for a comparison with the DGLAP approach at $\sqrt{s}=14$ TeV.

\begin{figure}[h]
\centerline{\begin{tabular}{cc}
\psfrag{dsigmaddeltaphi}{$\frac{1}{{\sigma}}\frac{d{\sigma}}{d \phi}$}
\psfrag{DPhi}{\scalebox{0.9}{$\Delta \phi$}}
\psfrag{deltaY6}{\hspace{-0.1cm}\scalebox{0.6}{$Y=6$}}
\psfrag{deltaY7}{\hspace{-0.1cm}\scalebox{0.6}{$Y=7$}}
\psfrag{deltaY8}{\hspace{-0.1cm}\scalebox{0.6}{$Y=8$}}
\hspace{-0.5cm} \includegraphics[width=5.5cm]{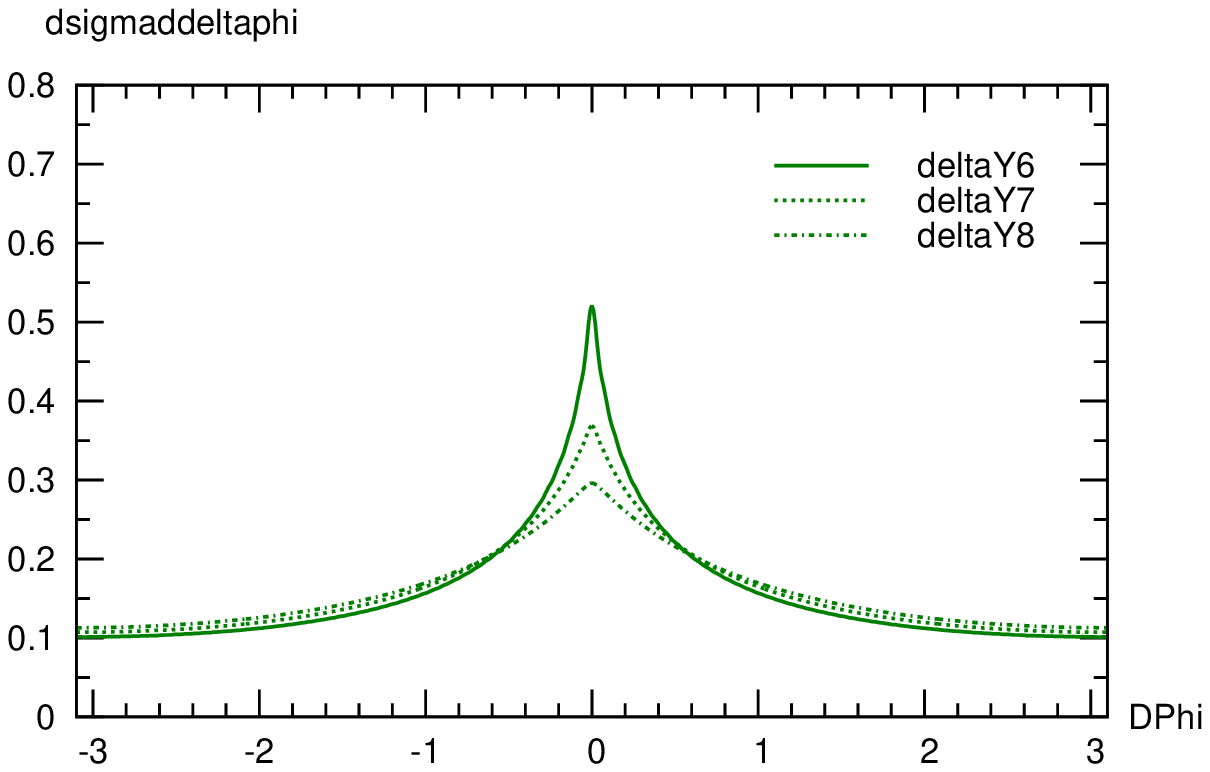} &
\psfrag{dsigmaddeltaphi}{$\frac{1}{{\sigma}}\frac{d{\sigma}}{d \phi}$}
\psfrag{DPhi}{\scalebox{0.9}{$\Delta \phi$}}
\psfrag{deltaY6}{\hspace{-0.1cm}\scalebox{0.6}{$Y=6$}}
\psfrag{deltaY7}{\hspace{-0.1cm}\scalebox{0.6}{$Y=7$}}
\psfrag{deltaY8}{\hspace{-0.1cm}\scalebox{0.6}{$Y=8$}}
\hspace{0.5cm}\includegraphics[width=5.5cm]{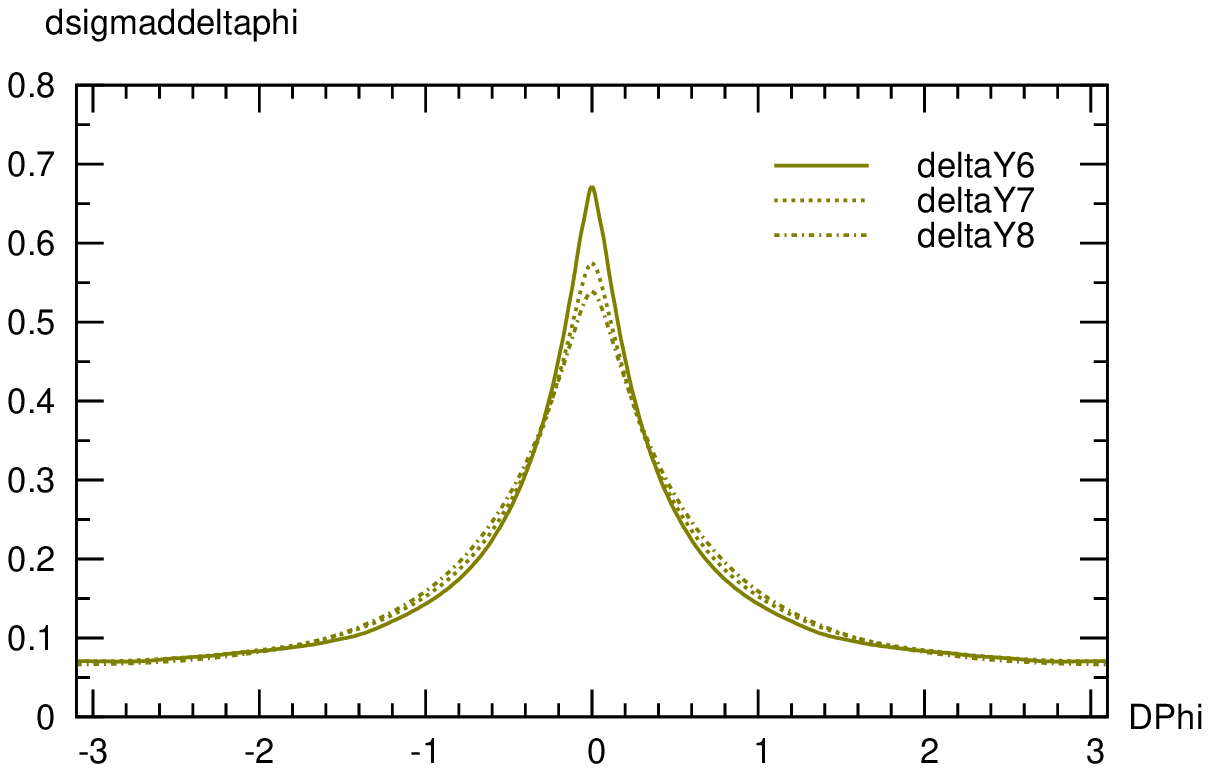} \\
\vspace{-0.1cm} & \vspace{-0.1cm} \\
\end{tabular}}
\caption{Azimuthal correlation. Left: NLL Green's function combined with LL vertices. Right:  NLL Green's function combined with NLL vertices.}
\label{Fig:Delta_phi_distribution}
\end{figure}

In order to get more insight into the azimuthal correlation between jets, we have studied the  
$\Delta \phi$ distribution, a quantity which is accessible at experiments like \textsc{ATLAS} and \textsc{CMS}.
Computing $\langle \cos(n \phi) \rangle$ up to large values of $n$ gives access to the angular distribution, since
\beq
\frac{1}{{\sigma}}\frac{d{\sigma}}{d \phi}
~=~ \frac{1}{2\pi}
\left\{1+2 \sum_{n=1}^\infty \cos{\left(n \phi\right)}
\left<\cos{\left( n \phi \right)}\right>\right\}\,.
\label{azimuthal_dist} 
\eq
The results, for various values of $Y$, are shown in Fig.~\ref{Fig:Delta_phi_distribution}.
Our 
full NLL treatment, when compared with a mixed NLL+LL approach, predicts 
less decorrelation for the same $Y$, and a 
slower decorrelation with increasing $Y$. 

\end{document}